\documentclass[11pt]{article}
\usepackage{amssymb,amsmath} 
\usepackage{graphicx}
\def\={\equiv} 
\def\div{\nabla\cdot } 
\def\pl{\partial}

\newcommand{\bib}{\bibitem}

\newcommand{\nt}{\notag}
\newcommand{\ci}{\cite}
\newcommand{\lab}{\label}

\newcommand{\eq}{\eqref}

\newcommand{\bx}[1]{ \boxed{#1}}

\newcommand{\lp}{\left(}
\newcommand{\rp}{ \right)}
\newcommand{\lb}{ \left[}
\newcommand{\rb}{\right]}

\newcommand{\LB}{\left\lbrace}
\newcommand{\RB}{\right\rbrace}

\newcommand{\0}[1]{{(#1)}}

\newcommand{\2}[1]{{\tilde #1}}
\newcommand{\3}[1]{{\boldsymbol #1}}

\newcommand{\bh}[1]{{\boldsymbol{\hat #1}}}

\newcommand{\6}[1]{_{\scriptscriptstyle#1}}

\newcommand{\8}{\infty}
\newcommand{\9}[1]{^{\,\scriptscriptstyle#1}}

\def\a{\alpha} 
\def\b{\beta} 

\def\c{\chi}
\def\d{\delta} 
\def\e{\varepsilon} 

\def\f{\phi}

\def\l{\lambda} 
\def\m{\mu} 
\def\n{\nu}
\def\o{\omega} 
\def\p{\pi} 

\def\q{\theta} 
 
\def\r{\rho}
 
\def\s{{\sigma}} 

\def\t{\tau}

\def\D{\Delta}

\def\O{\Omega} 

\def\Q{\Theta}

\def\Y{\Psi}

\newcommand{\hb}[1]{{\ \text{#1}\ }}

\newcommand{\db}{{\,{\rm d}\kern-.9ex {^-}}\!}
\newcommand{\dir}{{\pl\kern-1.2ex {/}}}

\newcommand{\app}{\approx} 
\newcommand{\cc}[1]{{{\mathbb C\hskip.5pt}^{#1}}}

\newcommand{\curl}{\nabla\times}

\newcommand{\grad}{\nabla} 

\newcommand{\ie}{{\it ie, }}

\newcommand{\im}{{\,\rm Im}\ }  
\newcommand{\imp}{\ \Rightarrow\ }
\newcommand{\inv}{^{-1}}

\newcommand{\ir}{\int_{-\infty}^\infty} 

\newcommand{\lra}{\leftrightarrow}

\newcommand{\plra}{\pl^{\kern-1.25ex^\lra}}
\newcommand{\qq}{\quad} 
\newcommand{\qqq}{\qquad} 
\newcommand{\re}{{\,\rm Re}\  }   
\newcommand{\rr}[1]{{{\mathbb R}^{#1}}}

\newcommand{\sgn}{{\,\rm Sgn \,}}
\newcommand{\sh}[1]{\hskip#1ex} 
\newcommand{\sr}{\sqrt}

\newcommand{\sv}[1]{\vskip#1ex}

\def\XXint#1#2#3{{\setbox0=\hbox{$#1{#2#3}{\int}$}
     \vcenter{\hbox{$#2#3$}}\kern-.5\wd0}}

\def\HB{\hfill\break}

\def\bib#1{\bibitem[#1]{#1}}
\begin{document}

\title{Making electromagnetic wavelets}

\author{Gerald Kaiser\thanks{
Research supported by AFOSR Grant \#F49620-01-1-0271.}\\
Center for Signals and Waves\\
kaiser@wavelets.com  $\bullet$\ www.wavelets.com}


\maketitle

\begin{abstract} 
\noindent Electromagnetic wavelets are constructed using  scalar wavelets as   superpotentials, together with an appropriate polarization. It is shown that \it oblate spheroidal antennas, \rm which are ideal for their production and reception, can be made by deforming and merging two branch cuts. This
determines a unique field on the interior of the spheroid which gives the boundary conditions for the surface charge-current density necessary to radiate the wavelets. These sources are computed, including the \it impulse response \rm of the antenna. 
\end{abstract}

\section{Complex distance and its branch cuts}

Electromagnetic wavelets were introduced in \ci{K94} as localized solutions of Maxwell's equations. They are `wavelets' in the historical sense of Huygens as well as the modern one: being generated from a single `mother wavelet' by conformal transformations including translations and scaling, they form \it frames \rm that provide analysis-synthesis schemes for general electromagnetic waves. Together with their  scalar (acoustic) counterparts, they have been called  \it physical wavelets. \rm It was pointed out that they can, in principle, be emitted and absorbed causally, and applications to radar and communications have been proposed \ci{K96, K97, K1} based on their remarkable ability to focus sharply and without sidelobes. Similar objects, long studied in the engineering literature under the name \it complex-source pulsed beams, \rm have been used to build beam summation methods and analyze the behavior of solutions (see \ci{HF1} for a recent review).  However, to implement the proposed applications to radar and communications, the wavelets must be \it realized \rm by simulating their \it sources. \rm This has proved to be difficult, and detailed investigations have only been made recently \ci{HLK0, K3}. Here I present a new and rather complete analysis of the sources based on an insight which I believe is a key to their realization: \it the sources can be constructed from the very branch cuts that give the wavelets their remarkable properties. \rm

Physical wavelets are based on the idea of displacing a point source to \it complex coordinates. \rm Since a \it real \rm translation gives nothing new, it suffices to discuss a point source with purely imaginary coordinates $i\3a$. It will be seen that this results in a \it real, coherent, extended source distribution \rm parameterized by the single vector $\3a$, much as an antenna dish can be described by a single vector giving the orientation and radius of the dish.  

Recall the definition of the \it complex distance \rm $\s$ from the imaginary source point $i\3a$ to the real field point $\3r$,
\begin{align}\lab{s}
\s(\3r-i\3a)&=\sr{(\3r-i\3a)\cdot(\3r-i\3a)}
=\sr{r^2-a^2-2i\3a\cdot\3r}.
\end{align}
For each fixed source location $i\3a\ne\30$, its branch points form a circle:  
\begin{align*}
\s=0\imp \3r\in \5C=\{\3r\in\rr3: r=a,\ \3a\cdot\3r=0\}.
\end{align*}
It is important to note the \it topological \rm difference between $\3a\ne \30$, when $\rr3-\5C$ is multiply connected, and $\3a=\30$, when $\5C$ contracts to the origin and $\rr3-\5C$ becomes simply connected.  Writing 
\begin{align}
\s=p-iq, 
\end{align}
note that \eq{s} implies
\begin{align*}
r^2-a^2=p^2-q^2,\qq  \3a\cdot\3r=pq,
\end{align*}
from which one easily obtains the relations to the cylindrical coordinates with $z$-axis parallel to $\3a$:
\begin{align}\lab{pq}
az=pq,\qq a\r=a\sr{r^2-z^2}=\sr{p^2+a^2}\sr{a^2-q^2}.
\end{align}
This gives an important bound on the `complexness' of $\s$ in terms of the `complexness' of its argument:  
\begin{align*}
\bx{|q|\le a, \hb{\ or\ } |\im \s(\3z)|\le |\im \3z|  ,\qq  \3z=\3r-i\3a.}
\end{align*}
It follows from \eq{pq} that 
\begin{align*}
\frac{\r^2}{a^2+p^2}+\frac{z^2}{p^2}=
\frac{\r^2}{a^2-q^2}-\frac{z^2}{q^2}=1,
\end{align*}
hence the level surfaces of $p^2$ form a family of  \it  oblate spheroids \rm
\begin{align}\lab{Sp}
\5S_p=\{\3r:  p^2={\rm const} >0\}
=\LB \3r: \frac{\r^2}{a^2+p^2}+\frac{z^2}{p^2}=1\RB
\end{align}
and those of $q^2$ form the orthogonal family of one-sheeted  \it  hyperboloids \rm
\begin{align}\lab{Hq}
\5H_q\equiv \{\3r: 0<q^2={\rm const}<a^2\}
=\LB\3r: \frac{\r^2}{a^2-q^2}-\frac{z^2}{q^2}=1\RB.
\end{align}
To complete the picture, we include the following \it degenerate \rm members of these families,
\begin{align}
\5S_0&=\{\3r: 0\le\r\le a,\ z=0\}\lab{S0}\\
\5H_0&=\{\3r: a\le\r<\8,\ z=0\}\lab{H0}\\
\5H_a&=\{\3r: \r=0,\ -\8<z<\8\}, \nt 
\end{align}
where $\5S_0$ and $\5H_0$ each form a  twofold cover  of the indicated sets.

\begin{figure}[ht]
\begin{center}
\includegraphics[width=3.5 in]{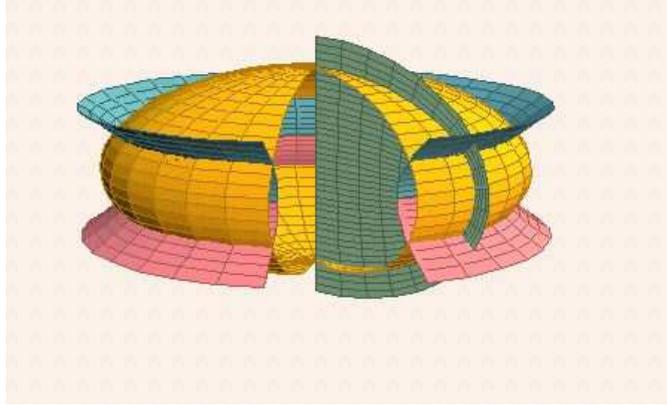}
\caption{The level surfaces of $\s$ form an oblate spheroidal coordinate system}
\label{zOSCS}
\end{center}
\end{figure}

The families $\5S_p$ and $\5H_q$ are depicted in Figure \ref{zOSCS}, along with the azimuthal half-plane $\f$=constant. They are  \it  confocal, \rm with the branch circle $\5C$ as their common focal set. Note that the intersection of $\5S_p$ with $\5H_q$ consists of two circles whose further intersection with the azimuthal half-plane consist of two points for each choice of $(p^2, q^2, \f)$. When $p=q=0$, the two circles merge with the branch circle $\5C$. The set of numbers
\begin{align*}
(\s,\f)\=(p,q,\f),\qq -\8<p<\8,\ -a\le q\le a,\ 0\le\f<2\p
\end{align*}
therefore gives a \it twofold cover \rm of $\rr3-\,\5C$. To obtain a coordinate system, we must choose between the two covers, and this amounts to choosing a  \it  branch cut \rm that makes $\s$ single-valued, as explained below. This will result in a one-to-one correspondence between $(\s,\f)$ and points $\3r\in\rr3$ not on the branch cut, giving an  \it  oblate spheroidal coordinate system. \rm

If we continue $\s$ analytically around a closed loop threading the branch circle $\5C$, it returns with its sign reversed. To make it single-valued, it is therefore necessary to choose a branch cut that prevents the completion of the loop. Note from \eq{s} that
\begin{align}\lab{far0}
r\gg a&\imp \s\app \pm(r-ia\cos\q).
\end{align}
The spatial region with $r\gg a$ will be called the  \it  far zone \rm (we need $a>0$ here to set the scale). Since we want $\s$ to generalize the usual positive distance $r$, we insist that 
\begin{align}\lab{far}
r\gg a\imp p\app r,\ q\app a\cos\q.
\end{align}
If follows from \eq{far} that the branch cut $\5B$  is  \it  bounded \rm since it must be entirely contained inside any spheroid $\5S_p$ with $p^2$ sufficiently large, and its boundary must be the branch circle:
\begin{align}\lab{dBC}
\pl\5B=\5C.
\end{align}
(The alternative is a branch cut extending from $\5C$ to infinity, but this violates \eq{far}.) $\5B$ is therefore a  \it  membrane spanning \rm  $\5C$, and   \it  any \rm such membrane will do. The situation is best understood topologically.  The analytic continuation of the distance has opened up a window connecting the two branches of $r=\sr{\3r\cdot\3r}$, thus making $\rr3-\,\5C$ multiply connected. The spherical coordinates $r$ and $\q$ merge analytically into $\s$, which is double-valued, and the choice of a branch cut $\5B$ makes $\rr3-\5B$ simply connected and $\s$ single-valued.  

Let $\s\60$ denote the complex distance with the flat disk $\5S_0$ as branch cut:
\begin{align}\lab{so}
\s\60=p\60-iq\60,\qq p\60\ge 0, \ -a\le q\60\le a.
\end{align}
The complex distance $\s\6{\5B}$ with $\5B$ as branch cut is now defined as follows.  Choose an arbitrary reference point $\3r_0$ in the far zone, where $\s\6{\5B}=\s\60$.  To find $\s\6{\5B}$ at any other point $\3r$, continue analytically along an arbitrary path from $\3r_0$ to $\3r$, with the following rule:  \it whenever the path crosses $\5B$, $\s\6{\5B}$ changes sign. \rm This gives a unique definition of $\s\6{\5B}$, and both $p\6{\5B}$ and $q\6{\5B}$ have a jump discontinuity across the interior of $\5B$. Of course, $\s\6{\5B}=\s\60=0$ on $\5C$.

A general branch cut can be specified by a function $\c(p,\f)$ as
\begin{align}
\5B=\{\3r\in\rr3:  p=\c(q,\f),\ -a\le q\le a,\ 0\le\f\le 2\p\},
\end{align}
where the  \it cut function \rm $\c$ must satisfy 
\begin{align}\lab{chi}
\c(-q,\f)=-\c(q,\f),\qq \c(q, 2\p)=\c(q,0).
\end{align}
The first condition ensures that $\s$ changes sign across $\5B$, while the second ensures that it is continuous across the half-plane $\f=0$ on each side of $\5B$. 
Note that $\5B$ need not be cylindrically symmetric. We will be especially interested in the cuts $\5B_\a$ defined by the cylindrical cut functions
\begin{align}\lab{qhat}
\c_\a(q)=\a\sgn\0q.
\end{align}
Note that on $\5B_\a$ we have
\begin{align*}
az=pq=\a q\sgn\0q=\a |q|.
\end{align*}
If $\a>0$, then the values $q\ne 0$ generate the \it upper spheroid: \rm 
\begin{align*}
\5S_\a\9+=\{\3r\in\5S_\a: z>0\}.
\end{align*}
But this does not include the branch circle $\5C$, so the bounding condition \eq{dBC} is not satisfied. The problem is that $\sgn\0q$ is undefined at $q=0$, and the set of all points with $q=0$ is the  degenerate hyperboloid $\5H_0$ \eq{H0}. Hence we \it define \rm the part of $\5B_\a$ with $q=0$ as the \it apron \rm bridging the gap between $\5C$ and $\5S_\a\9+$,
\begin{align*}
\5B_\a\90=\{\3r: a\le\r\le \sr{a^2+\a^2},\ z=0\}. 
\end{align*}
Thus, for $\a>0$, $\c_\a\0q$ defines  the  \it upper spheroidal branch cut: \rm
\begin{align}\lab{Bp}
\5B_\a=\5S_\a\9+\cup\5B_\a\90\,,\qq \pl\5B_\a=\5C.
\end{align}
Similarly,  $\c_{-\a}\0q$ defines the \it lower spheroidal cut: \rm 
\begin{align}\lab{Bm}
\5B_{-\a}=\5S_\a\9-\cup\5B_\a\90\,,\qq \pl\5B_{-\a}=\5C.
\end{align}
As $\a\to 0$, both cuts contract to  the doubly covered flat disk spanning $\5C$, which is the degenerate spheroid $\5S_0$ \eq{S0}.

\begin{figure}[ht]
\begin{center}
\includegraphics[width=4 cm]{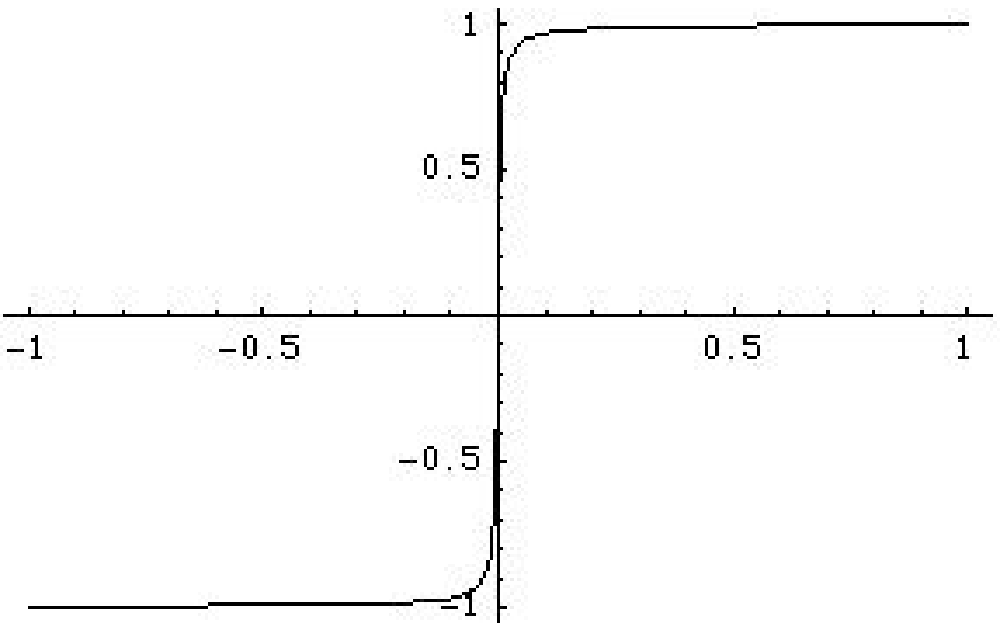}\sh2
\includegraphics[width=6 cm]{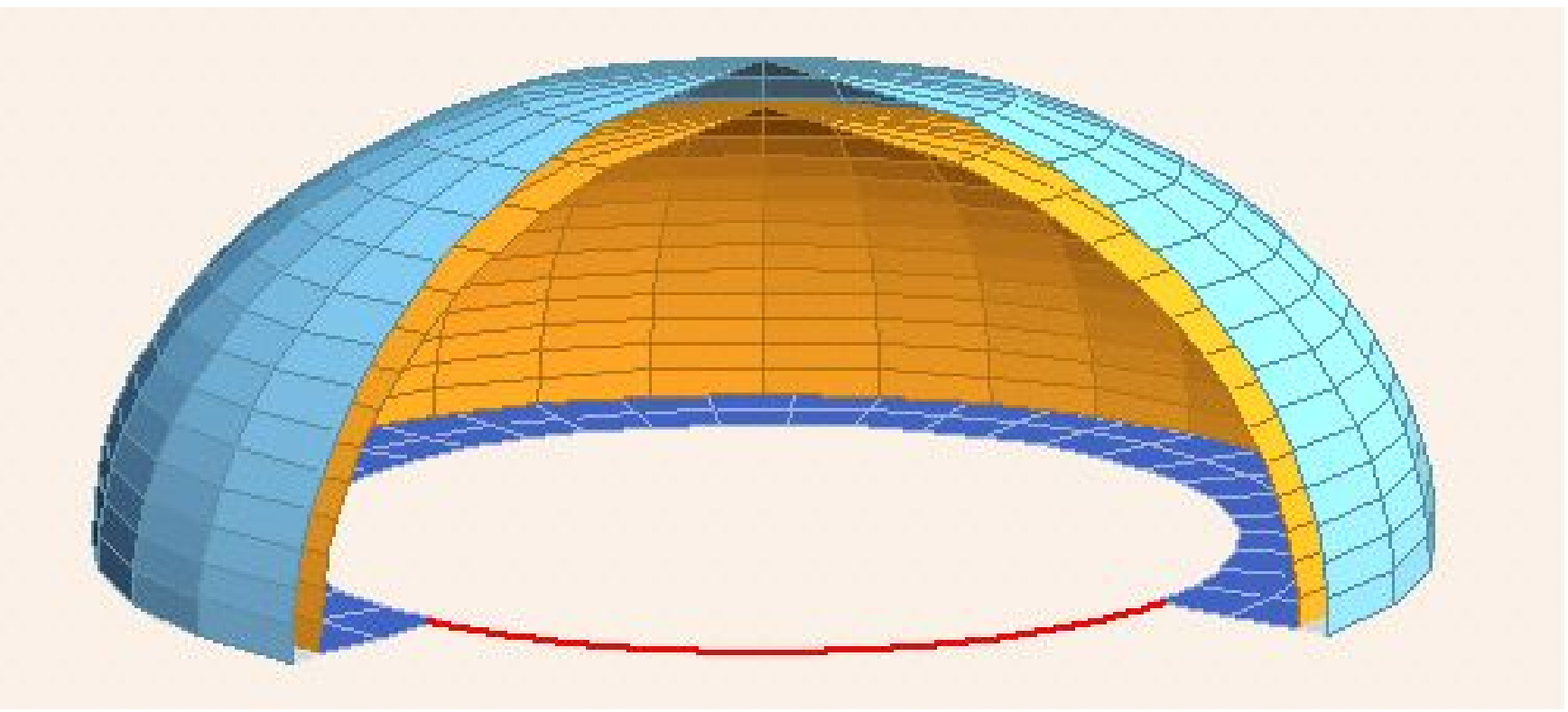}
\caption{The cut function $\c\0q=X_\e\0q$ with $\e=.005$  and its branch cut with $a=1$.  The two sheets have been purposely separated to show the double cover.}
\label{zDouble}
\end{center}
\end{figure}

\it Every \rm branch cut $\5B$ is doubly covered.  Consider any simply connected, closed surface $\5S$ containing $\5C$ in its interior. Think of $\5S$ as a balloon and of  $\5C$ as a rigid wire ring. Now \it deflate \rm the balloon, and you have a branch cut bounded by $\5C$. In particular, if we take $\5S=\5S_\a$ and keep the upper spheroid rigid while deflating, the balloon stretches around the ring to cover the underside of $\5S_\a\9+$ and we obtain the cut \eq{Bp} (see Figure \ref{zDouble}).

The image of a branch cut as a balloon enclosing the singular ring is similar  to Penrose's idea of \it cosmic censorship \rm  in general relativity \ci[Chapter 5]{W99}, where a \it horizon \rm ($\5S$) prevents the outside observer from seeing a \it naked singularity \rm ($\5C$). In light of the connection with Newman's analytic Coulomb field (see the discussion below \eq{Coul}), the two may in fact be closely related.

The discontinuity of $\c_\a$ in \eq{qhat} causes two problems: the $q=0$ contribution is undefined (hence the aprons $\5B_\a\90$ had to be chosen `by hand'), and  the resulting cut had a sharp edge. For computational purposes, it may be better to use smooth cut functions to avoid both problems.  Let $\e>0$ and define 
\begin{align}\lab{sigma}
X_\e\0q=\frac1\p\im\ln\lp\frac{\e+iq}{\e-iq}\rp,\qq \c(q)=\a X_\e\0q.
\end{align}
For $\e\ll a$, $X_\e\0q$ is a smoothed version of $\sgn\0q$ and the resulting branch cut closely approximates $\5B_\a$ without the need to define the apron separately. This is shown in Figure \ref{zDouble}.

\section{Scalar wavelets}

For any fixed choice of branch cut $\5B$, we now denote the complex distance simply by $\s$. Scalar wavelets are then defined as the \it  retarded \rm solutions
\begin{align}\lab{Yr}
\Y(\3z,\t)&=\frac{g(\t-\s)}{\s}\equiv \frac{g_r}{\s}\,, \ 
\3z=\3r-i\3a,\ \t=t-ib,
\end{align}
where we have set the propagation speed $c=1$ (otherwise $g_r=g(\t-\s/c)$) and $g$ is the  \it  analytic-signal transform \rm of a driving signal $g_0\0t$, 
defined\footnote{This is a special case of a multidimensional definition; see \ci{K3}.}
as the convolution of $g_0$ with the Cauchy kernel:
\begin{align}
g\0\t&=\frac1{2i\p}\ir\frac{g_0(t')\,dt'}{\t-t'}\,,\qqq\qq \t=t-ib\nt\\
&=\frac b{2\p}\ir\frac{g_0(t')\,dt'}{(t'-t)^2+b^2}
+\frac i{2\p}\ir\frac{(t'-t)g_0(t')\,dt'}{(t'-t)^2+b^2}\nt\\
&= g_1(t,b)+ig_2(t,b).\lab{g}
\end{align}
$g_1(t,b)$ and $g_2(t,b)$ are smoothed versions of $\tfrac12 g_0\0t$ and its  \it  Hilbert transform, \rm with $b$ as the smoothing parameter.  We assume that $g_0\0t$ decays at infinity, from which it follows that $g\0\t$ is analytic in the upper and lower complex time half-planes $\4C\9\pm$. The original driving signal can be recovered as the boundary value
\begin{align*}
g_0\0t=\lim_{b\to+0}\lb g(t-ib)-g(t+ib)\rb.
\end{align*}
(The limit of the \it sum \rm gives the Hilbert transform.)
In particular, if $g_0$ vanishes on \it any \rm open interval $I$, this interval becomes a window between the upper and lower half-planes through which the
functions $g(t\pm ib)$ can be connected so that they are both part of a \it single \rm analytic function. (This is a special case of the \it edge of the wedge theorem \rm in higher dimensions; see \ci{K3}.) Since every practical driving signal vanishes at least in the remote past, this property will be assumed. Note that this excludes time-harmonic driving signals, which are however idealizations.

Now consider the numerator of \eq{Yr}, 
\begin{align*}
g(\t-\s)=g(t-p-i(b-q)).
\end{align*}
Suppose that $|b|\le a$. Then $g(\t-\s)$ is  undefined  along the semi-hyperboloid where $q(\3r)=b$, \it  except \rm when $t-p(\3r)$ is in the zero-set of $g_0$. On the other hand, if $|b|>a$, then $b-q(\3r)$ vanishes nowhere and $g(\t-\s)$ is analytic at all $(\3r,t)$.
Therefore we assume from now on that
\begin{align}\lab{ba}
\bx{\ |b|>a\ }
\end{align}
so that $g(\t-\s)$ is defined unambiguously everywhere. The imaginary source coordinates must therefore belong either to the  \it   future cone \rm or to the  \it  past cone \rm of space-time,
\begin{align*}
b>a&\imp (\3a, b)\in V\6+\\
b<-a &\imp (\3a, b)\in V\6-\,,
\end{align*}
which means that the complex 4-vector from the source point $iy=i(\3a, b)$ to the field point $x=(\3r, t)$ belongs either to the \it forward tube \rm or the \it backward tube \rm of complex space-time \ci{SW64, K3}, 
\begin{align}\lab{Tpm}
\5T\6\pm=\{z=x-iy\in\cc4: y\in V\6\pm\},\qq  x=(\3r, t),\  y=(\3a, b).
\end{align}

The source distribution of $\Y$ is now \it defined \rm as a generalized function by applying the wave operator,
\begin{align}\lab{S}
S\0z=(\pl_t^2-\grad^2)\Y\0z=\square_x\Y\0z,
\end{align}
where $\Box_x$ indicates that the operator acts only on the real space-time variables $x$ of the field point. It is well known that 
\begin{align}\lab{hr}
\Box\,\frac{h(t-r)}r=4\p h\0t\d(\3r)  
\end{align}
for any differentiable function $h$, and this can be extended to $\Y\0z$. Since $\Y$ is differentiable in $\3r$ everywhere outside of the branch cut $\5B$, \eq{hr} suggests that $S$ is a  (Schwartz) \it  distribution supported on $\5B$, \rm a conclusion borne out by a rigorous analysis \ci{K3}. The discontinuity of $\Y$ across $\5B$ gives a combination of \it simple and double layer \rm terms of $S$ on $\5B$ \ci{K3}.

The frequency content of $g\0\t$ determines that of $\Y$ and should therefore be understood. Substituting the Fourier representation of $g_0$ into the definition \eq{g} and reversing the order of integration gives
\begin{align}
g\0\t&=\frac1{2i\p}\ir\frac{dt'}{\t-t'}\ir\frac{d\o}{2\p}\ \1g_0\0\o e^{-i\o t'}\nt\\
&=\ir\frac{d\o}{2\p}\ \1g_0\0\o\,\frac1{2i\p}\ir\frac{e^{-i\o t'}\,dt'}{\t-t'}.\lab{g2}
\end{align}
The contour in the second integral can be closed in the lower half-plane if $b>0$ and in the upper half-plane if $b<0$, giving 
\begin{align}\lab{ast}
g\0\t=\sgn\0b\ir\frac{d\o}{2\p}\ \1g_0\0\o\,\Q(\o b) e^{-i\o\t},\qq \t=t-ib,
\end{align}
where $\Q(\o b)$ is the Heaviside step function.
Thus if $b>0$, $g$ contains only the positive-frequency components of $g_0$, and
 if $b<0$, it contains only the negative-frequency components. In either case, the factor $e^{-\o b}$ in the extended Fourier kernel
 \begin{align*}
e^{-i\o\t}= e^{-\o b} e^{-i\o t}
\end{align*} 
acts as a \it low-pass filter, \rm substantially damping frequencies $|\o |\gg b\inv$ and 
thus  smoothing out $g\0\t$. If the driving signal $g_0$ is assumed \it real, \rm then $\1g_0(\pm\o)$ are related by complex conjugation and therefore so are $g(t\mp ib)$. If 
$g_0$ is complex, then $\1g_0(\pm\o)$ are unrelated and so are $g(t\mp ib)$.

\bf Example: \rm Let $g\0\t$ be the $(n-1)$-st derivative of the Cauchy 
kernel,\footnote{The driving signal is the singular distribution 
$g_0\0t=(i\pl_t)^{n-1}\d\0t$, but this can be approximated.}
\begin{align}\lab{C}
g(\t)=C_n\0\t=(i\pl_t)^{n-1} \frac1{2i\p\t} 
=\frac{(n-1)!}{2\p i^n\t^n},
\end{align}
whose Fourier transform is
\begin{align}\lab{Cn}
\1C_n(\o, b)=\ir dt\ e^{i\o t} C_n(t-ib)=\sgn\0b\Q(\o b)\,\o^{n-1}e^{-\o b}.
\end{align}
Thus, while $b$ acts to suppress high frequencies, $n>1$ acts to suppress low frequencies and we end up with a \it band-pass filter \rm whose effective center frequency and bandwidth are given by a Poisson distribution,
\begin{align}\lab{on}
\o_n=\frac n b,\qqq \D\o=\frac{\sr{n}}{| b |}.
\end{align}
The behavior of $\Y$ in the far zone is governed by that of $g(\t-s)$. By \eq{far}, 
\begin{align*}
C_n(\t-\s)=\frac{(n-1)!}{2\p i^n(\t-\s)^n}
\app\frac{(n-1)!}{2\p [(b-a\cos\q)+i(t-r)]^n}
\end{align*}
is a \it pulse \rm with angle-dependent duration 
\begin{align}\lab{T}
T\0\q=|b-a\cos\q|\ge |b|-a=T_{\rm min}>0,
\end{align}
being shortest at $\q=0$ if $b>a$ and at $\q=\p$ if $b<-a$.
While the pulse duration is independent of $n$, the \it strength \rm of the peak depends jointly on the size of $n$ and the smallness of $b-a$:
\begin{align}\lab{M}
M\0\q=|g(\t-\s)\!\bigm|_{t=r}\app\frac{(n-1)!}{2\p T\0\q^n}.
\end{align}
To get a measure of the diffraction angle, assume $b>a$ for definiteness. Fix $\b>0$ and look for the angle $\q_\b$ at which
\begin{align*}
M(\q_\b)= e^{-\b}M\00.
\end{align*}
Then
\begin{align*}
(b-a\cos\q_\b)^n=e^\b(b-a)^n,
\end{align*}
which gives
\begin{align}\lab{diff}
2\sin^2(\q_\b/2)=1-\cos\q_\b=(e^{\b/n}-1)\frac{b-a}{a}\,.
\end{align}
Thus, $\q_\b$ can be made small either by choosing $b-a\ll a$, or  $n\gg\b$. In either case, the right side gives $\q_\b^2/2$. A reasonable measure is obtained with $\b=1$.

\section{From scalar to vector wavelets}

It is well known that every electromagnetic field can be derived from a pair of real scalar potentials, the most well-known examples of which are the \it Whittaker \rm and \it Debye \rm superpotentials \ci{N55}. In this section we use the scalar wavelet $\Y$ as a \it complex \rm Whittaker superpotential. Although this is equivalent to using a pair of real potentials, disentangling the real and imaginary parts leads to unecessarily complicated expressions, something like taking the real and imaginary parts of a complicated analytic function $f(x+iy)$ in order to obtain two real harmonic functions. To see how bad it gets, note from \eq{Flong} that the fields and currents contain terms of the type $\s^{k-3}g^{(k)}(\t-\s)$ with $k=0, 1,2$. In the \it simplest \rm case $k=2$ (which will give the radiation terms of the field), \eq{g} gives
\begin{align}
\frac{\ddot g(\t-\s)}{\s}&=\frac{g_1(t-b, b-q)+ig_2(t-b, b-q)}{p-iq}\nt\\
\re\LB \frac{\ddot g(\t-\s)}{\s}\RB&=
\frac{p\ddot g_1(t-b, b-q)-q\ddot g_2(t-b, b-q)}{p^2+q^2}\nt\\
\im\LB\frac{\ddot g(\t-\s)}{\s}\RB
&=\frac{p\ddot g_2(t-b, b-q)+q\ddot g_1(t-b, b-q)}{p^2+q^2},\lab{real}
\end{align}
and it is clear that the real expressions quickly become unmanageable.  Thus, although we work with complex potentials and fields, we view this as a very compact and efficient way of computing the \it real \rm fields. In particular, our expressions contain \it nothing extraneous \rm since their imaginary as well as real parts have a direct physical significance. This strategy is based on the \it analyticity \rm of $\Y$ outside the source region, which will indeed make \it  harmonic pairs \rm out the fields $\3D$ and $\3B$, as seen below.

With $\Y$ as a complex Whittaker superpotential, we define the retarded complex \it  Hertz potential \rm 
\begin{align}\lab{Z}
\bx{\3Z=\Y\3\p}
\end{align}
where $\3\p$ is a fixed complex \it polarization vector \rm that can be seen \ci{K3} to be a combination of (real)  \it electric and magnetic dipole moments. \rm The real and imaginary parts of $\3Z$
\begin{align*}
\3Z=\3Z_e+i\3Z_m
\end{align*}
are interpreted as \it electric and magnetic Hertz vectors \rm \ci[pp 84--85]{BW99}. They generate a 4-vector potential $A_\m\  (\m=0, 1,2,3)$ by
\begin{align}\lab{A}
A_0=-\div\3Z_e,\qq \3A=\pl_t\3Z_e+\curl\3Z_m
\end{align}
which automatically satisfies the Lorenz condition
\begin{align}\lab{lor}
\pl_tA_0+ \div\3A=0.
\end{align}
In turn, it follows from potential theory (or the Poincar\`e lemma for differential forms) that \it every \rm  4-vector potential satisfying \eq{lor} can be written in the form \eq{A}, so this representation is quite general. (We can even dispense with the Lorenz condition by performing a gauge transformation on $A_\m$. See \ci{N55} for an excellent and thorough account of Hertz potentials and their enormous gauge group.) The real vector fields $\3P_e$ and $\3P_m$ defined by
\begin{align}\lab{P}
\3P=\3P_e+i\3P_m=\Box\3Z=(\Box\Y)\3\p=S\3\p
\end{align}
are the \it electric and magnetic polarization densities. \rm 
They are distributions supported spatially on the branch cut $\5B$. Since we are in Lorenz gauge, the charge-current density is $J_\m=\Box A_\m$, hence 
\begin{align}\lab{J}
J_0=-\div\3P_e\,, \qq \3J=\pl_t\3P_e+\curl\3P_m\,,
\end{align}
with charge conservation guaranteed by the Lorenz condition. The polarization densities thus act as `potentials' for the charge-current density, a property inherited directly from \eq{A}.

The reason why Hertz potentials will be so useful can be seen by computing the fields:
\begin{align}\lab{B}
\3B=\curl\3A=\curl\curl\3Z_m+\pl_t\curl\3Z_e
\end{align}
and
\begin{align}
\3E&=-\grad A_0-\pl_t\3A=\grad\div\3Z_e-\pl_t^2\3Z_e-\pl_t\curl\3Z_m\nt\\
&=-\Box\3Z_e+\curl\curl\3Z_e-\pl_t\curl\3Z_m.\lab{E}
\end{align}
Taking into account \eq{P} gives
\begin{align}\lab{D}
\3D=\3E+\3P_e=\curl\curl\3Z_e-\pl_t\curl\3Z_m
\end{align}
which is a kind of `harmonic conjugate' of \eq{B}, so the \it real \rm fields $\3D, \3B$ can be expressed compactly in the \it complex \rm form \ci[pp 32--34]{S41}
\begin{align}\lab{F}\bx{
\3F\equiv \3D+i\3B=\curl\curl\3Z+i\pl_t\3Z.}
\end{align}
Note that outside the branch cut $\5B$, $\3P_e=\30$ and $\3D=\3E$. The Hertz formalism thus automatically takes account of the polarization, so that the expression \eq{D}, if interpreted as a distribution, is valid even within a singular source region. 

Again I emphasize that  $\3F=\3D+i\3B$ is ``real'' in the sense that $\3D$ and $\3B$ are real, physical fields. Yet $\3F$, like $\Y$, is analytic in the source-free complex space-time region \rm
\begin{align}\lab{TB}
\5T\6{\5B}=\{(\3r-i\3a,t-ib)\in\cc4:  |b|>a,\ \3r\notin\5B\}.
\end{align}
More simply, because of their spheroidal symmetry, $\Y$ and $\3F$ are analytic functions of the two complex variables $(\s,\t)$ in the region
\begin{align}\lab{UB}
\5U\6{\5B}=\{(\s,\t)\in\cc2:  |b|>a,\  p\ne \c\6{\5B}\0q\},
\end{align}
where $\c\6{\5B}\0q$ is the cut function for $\5B$. 
Thus $\3D$ and $\3B$ really are \it harmonic conjugates \rm as suggested earlier.  On the  other hand, $\3P$ and $J_\m$ characterize the \it singularities \rm spoiling analyticity in the source region, including the branch points and branch cuts.  This differs from the usual practice in the frequency domain,  where $(\3D, \3B)$ are the real parts of separate complex fields $(\3D_c\,, \3B_c)$, and  it might appear that these two representations are in conflict since the real fields cannot be extracted by taking the real and imaginary parts of  $\3D_c+i\3B_c$. To clarify this, consider the frequency components of $\3F$, 
\begin{align*}
\3F_\o=\ir dt\ e^{i\o t}\3F=\3D_\o+i\3B_\o,
\end{align*}
and note that since $\3F$ is complex, its positive-and negative-frequency components are \it independent. \rm Therefore a general monochromatic field consists of \it two \rm terms,
\begin{align*}
\3G_{\rm mono}&=e^{-i\o t}\3F_\o+e^{i\o t}\3F_{-\o}, \qq \o>0\\
&=e^{-i\o t}(\3D_\o+i\3B_\o)+e^{i\o t}(\3D_{-\o}+i\3B_{-\o})\\
&=e^{-i\o t}(\3D_\o+i\3B_\o)+e^{i\o t}(\3D^*_\o+i\3B^*_\o),
\end{align*}
where the reality conditions have been used on $\3D_\o$ and $\3B_\o$.
The representation of a monochromatic field is therefore no different from that of a general field:
\begin{align*}
\3G_{\rm mono}=\3D_{\rm mono}+i\3B_{\rm mono}
\end{align*}
with both fields \it real: \rm
\begin{align*}
\3D_{\rm mono}\0t&=2\re\!\LB e^{-i\o t}\3D_\o\RB\\
\3B_{\rm mono}\0t&=2\re \!\LB e^{-i\o t}\3B_\o\RB.
\end{align*}
Recall that for $b>a$ and $b<-a$, the analytic signal $g\0\t$ contains only positive- and negative-frequency components. Therefore
\begin{align*}
b>a&\imp \3D_\o+i\3B_\o=\30\  \forall \o<0\\
b<-a&\imp \3D_\o+i\3B_\o=\30\  \forall \o>0.
\end{align*} 
This shows that the monochromatic components of the electromagnetic wavelets satisfy
\begin{align*}
\3B_{\rm mono}\0t=2\re \!\LB ie^{-i\o t}\3D_\o\RB
=\3D_{\rm mono}(t-\p/2\o),
\end{align*} 
so $\3B$ trails $\3D$ if $b>a$, and it leads $\3D$ if $b<-a$. (Recall also that the pulse travels along $\pm\3a$ if $\pm b>a$.)  More generally, the wavelets are \it helicity eigenstates \rm with helicity $1$ if $b>a$ and $-1$ if $b<-a$. This concept applies not only the time-harmonic components but also to general time domain fields \ci{K3a}. As already mentioned, using the analytic combinations of fields also has the great advantage of compactness and simplicity over the alternative of disentangling the real and imaginary parts. 

\begin{figure}[ht]
\begin{center}
\includegraphics[width=3.5 in]{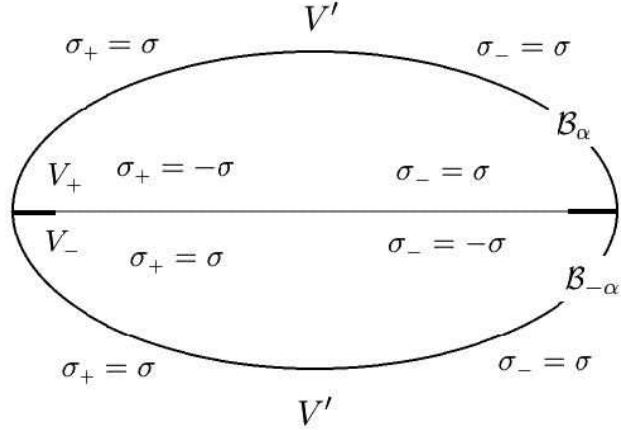}
\caption{Values of the branches $\s\6\pm$ of the complex distance function determined by the branch cuts $\5B_{\pm\a}$, given in terms of the branch $\s$ determined by the disk $\5S_0$.}
\label{zEllipseSigma}
\end{center}
\end{figure}

Before launching into the field computations, I want to prepare the way for computing \it equivalent currents \rm on the spheroid $\5S_\a$ in the coming section. Let us construct this spheroid from the two branch cuts $\5B_{\pm\a}$  given in \eq{Bp} and \eq{Bm}. Let us now denote by $\s$ the complex distance with the disk $\5S_0$ as branch  cut. This will be used as a reference for defining the complex distance functions with $\5B_{\pm\a}$ as cuts, which we denote by $\s\6\pm$. Let $V\6\pm$ be the volumes bounded by $\5S_\a\9\pm$ together with $\5S_0$, so that
\begin{align*}
\pl V\6+=\5S_\a\9+-\5S_0,\qq \pl V\6-=\5S_0-\5S_\a\9-,
\end{align*}
where the signs are related to the \it orientations \rm of $\5S_\a\9\pm$  and $\5S_0$ by $\3a$. The union and compliments of $V\6\pm$ will be denoted by
\begin{align*}
V=V\6+\cup V\6-\,,\qq 
V\6\pm'=\rr3-V\6\pm\,,\qq
V'&=\rr3-V.
\end{align*}
Now recall the  rule for crossing a branch cut \it other than the reference \rm cut $\5S_0$: $\s\6{\5B}$ changes sign. Thus, denoting the complex distance functions with respect to $\5B_{\pm\a}$ by $\s\6\pm$, we have
\begin{align}\lab{spm}
\s\6\pm= \begin{cases}
\s &\text{in\ } V\6\pm'\\ -\s&\text{in\ } V\6\pm
\end{cases}
\end{align}
as shown in Figure \ref{zEllipseSigma}.

\begin{figure}[ht]
\begin{center}
\includegraphics[width=3.5 in]{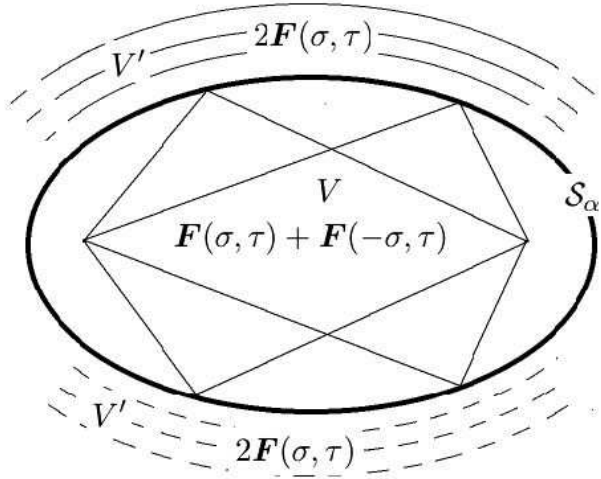}
\caption{Interior and exterior fields radiated by the oblate spheroid $\5S_\a$, represented as a combination of the two branch cuts $\5B_{\pm\a}$.}
\label{zEllipseField}
\end{center}
\end{figure}

The field radiated \it  jointly \rm by the two branch cuts $\5B_{\pm\a}$ is therefore
\begin{align*}
\3F_\a(\s,\t)= \begin{cases}
2\3F(\s,\t) & \text{in\ } V'\\ \3F(-\s,\t)+\3F(\s,\t) &\text{in\ } V\6+\\
\3F(\s,\t)+\3F(-\s,\t) &\text{in\ } V\6-\,.
\end{cases}
\end{align*}
Observe that  \it there is no field discontinuity in going from $V\6+$ to $V\6-$, hence \rm 
\begin{align}\lab{joint}
\3F_\a(\s,\t)= \begin{cases}
2\3F(\s,\t) & \text{in\ } V'\\ \3F(-\s,\t)+\3F(\s,\t) &\text{in\ } V
\end{cases}
\end{align}
as depicted in Figure \ref{zEllipseField}.
\sv2

The transition $\s\to-\s$ across a branch cut turns \it retarded \rm fields into \it advanced \rm fields since
\begin{align}\lab{crossing}
\Y(-\s,\t)=-\frac{g(\t+\s)}{\s}.
\end{align}
Although advanced fields are usually associated with acausal behavior, there is a perfectly causal explanation for \eq{crossing}. Consider the field radiated \it backward \rm from $\5B_\a$, as observed in $V\6+$. Due to the curvature of the back side of $\5B_\a$, this field \it converges \rm toward  the the focal ring  $\5C$ and, having passed though, it is no longer in $V\6+$ and therefore diverges normally. A similar argument explains why the field emitted \it forward \rm from $\5B_{-\a}$ first converges toward $\5C$ and then diverges away from it.  The usual acausal behavior associated with advanced fields is due the the assumption that they \it remain \rm advanced for the indefinite future. (This argument also applies to \it time-reversed acoustics \rm \ci{F0}, where time reversal occurs only in a bounded space-time region.)

It was shown in \ci{K3} that the sources of $\Y(\s,\t)$ and $\Y(-\s,\t)$ are equal and opposite; that is,  they form a \it source-sink pair: \rm
\begin{align}\lab{sink}
\Box \Y(-\s,\t)=-\Box\Y(\s,\t)=-S.
\end{align}
The proof is trivial for real point sources, where
\begin{align*}
\Box \,\frac{g_0(t\pm r)}r=-4\p g_0(t\pm r)\d(\3r)=-4\p g_0(t)\d(\3r).
\end{align*}
But it is more subtle for complex point sources because the \it extended delta function \rm
\begin{align*}
\2\d(\3z)=-\grad^2\,\frac1{4\p\s}\,,\qq \3z=\3r-i\3a
\end{align*}
with $\3a\ne\30$ fixed, is not supported at a single point but on the entire branch cut $\5B$ and thererfore 
\begin{align*}
f\0\s\2\d(\3z)\ne f\00\2\d(\3z).
\end{align*}
In fact, the left side is not even \sl defined \rm since $\s$ is discontinuous precisely on the disk supporting $\2\d(\3z)$; therefore, some care must be used in proving \eq{sink}. 

Equation \eq{sink} shows that the \it interior superpotential  $\Y(\s,\t)+\Y(-\s,\t)$ is sourceless,   as are the Hertz potentials and electromagnetic fields derived from it. \rm The interior field is 
\begin{align}\lab{Fo}
\bx{\ \3F_0(\s,\t)=\3F(\s,\t)+\3F(-\s,\t). \ }
\end{align}
Let us first compute the exterior field $\3F$, which will give the interior field by symmetrizing with respect to $\s$. Let
\begin{align*}
&\Y'\equiv \pl_\s\Y=-\frac{\dot g_r}{\s}-\frac{g_r}{\s^2},
\qq\dot g_r=\pl_t g_r(\t-\s)\\
&\Y''=\pl_\s\Y'=\frac{\ddot g_r}{\s}+\frac{2\dot g_r}{\s^2}+\frac{2g_r}{\s^3}
=\frac{\ddot g_r-2\Y'}{\s}\\
&\3u=\grad \s=\grad p-i\grad q=\frac{\3z}{\s}
\end{align*}
and note that $\3u$ is a complex unit vector:
\begin{align*}
\3u\cdot\3u=\frac{\3z\cdot\3z}{\s^2}=1.
\end{align*}
Thus
\begin{align*}
\curl\3Z&=\grad\Y\times\3\p=\Y'\3u\times\3\p\\
\curl\curl\3Z&=\Y''\3u\times(\3u\times\3\p)+\Y'\curl(\3u\times\3\p),
\end{align*}
and by a simple computation,
\begin{align*}
&\3u\times(\3u\times\3\p)=\l\3u-\3\p,\qq \l=\3u\cdot\3\p\\
&\curl(\3u\times\3\p)=-\frac{\l\3u+\3\p}{\s}\,.
\end{align*}
Therefore \eq{F} gives
\begin{align*}
\3F=\Y''(\l\3u-\3\p)-\frac{\Y'}{\s}(\l\3u+\3\p)+i\dot\Y'\3u\times\3\p
\end{align*}
or
\begin{align}\lab{Flong}
\3F=\lp \frac{\ddot g_r}{\s}+\frac{3\dot g_r}{\s^2}+\frac{3g_r}{\s^3}\rp\l\3u
-\lp\frac{\ddot g_r}{\s}+\frac{\dot g_r}{\s^2}+\frac{g_r}{\s^3}\rp\3\p
-i\lp \frac{\ddot g_r}{\s}+\frac{3\dot g_r}{\s^2}\rp\3u\times\3\p.
\end{align}
This expression will be written compactly as
\begin{align}\lab{F2}\bx{
\3F=L\l\3u-M\3\p-iN\3u\times\3\p\ }
\end{align}
where
\begin{align}
L&=\frac{\ddot g_r}{\s}+\frac{3\dot g_r}{\s^2}+\frac{3g_r}{\s^3}\nt\\
M&=\frac{\ddot g_r}{\s}+\frac{\dot g_r}{\s^2}+\frac{g_r}{\s^3}\nt\\
N&=\frac{\ddot g_r}{\s}+\frac{\dot g_r}{\s^2}\,.\lab{LMN}
\end{align}

We now examine the far field to see under what conditions the polarization vector $\3\p$ gives the strongest beams.  In the far zone \eq{far} we have
\begin{align*}
r\gg a\imp \s\app r-ia\cos\q,\qq \3u\app\3e_r.
\end{align*}
Therefore
\begin{align}
\3F_{\rm far}&=\frac{\ddot g(\t-\s)}r\lp \l\3e_r-\3\p-i\3e_r\times\3\p\rp\nt\\
&=-\frac{\ddot g(\t-\s)}r\lp \3\p\6\perp+i\3e_r\times\3\p\6\perp\rp,\lab{far2}
\end{align}
where
\begin{align*}
\3\p\6\perp=\3\p-(\3\p\cdot\3e_r)\3e_r
\end{align*}
is the component of $\3\p$ orthogonal to $\3r$ which, as expected, is the only one that matters in the far zone.  Note that while we have replaced $\s$ by $r$ in the denominator of \eq{far2}, the presence of $\im \s\app -r\cos\q$ in $g(\t-\s)$ plays an \it essential \rm role in determining both the collimation of the beam and the duration of the pulse, as already seen in \eq{M} for $g\0\t=C_n\0\t$.

The far field satisfies the \it helicity condition \rm
\begin{align}\lab{tem}
i\3e_r\times \3F_{\rm far}=\3F_{\rm far}\,,
\end{align}
or equivalently
\begin{align*}
\3B_{\rm far}=\3e_r\times\3D_{\rm far},\qq
\3D_{\rm far}=-\3e_r\times\3B_{\rm far}\,.
\end{align*} 
As we are interested mainly in the paraxial region of the far zone, \it the most efficient choice of $\3\p$ is orthogonal to $\3a$. \rm  Since
\begin{align*}
\3r=\3\r+z\bh a\imp
\3u=\frac{\3r-i\3a}{\s}=\frac{\3\r}{\s}+\frac{z-ia}{\s}\,\bh a,
\end{align*}
this implies 
\begin{align*}
\l=\3u\cdot\3\p=\frac{\3\r\cdot\3\p}{\s}.
\end{align*}
The far-zone energy density is
\begin{align*}
\5E_{\rm far}=\frac12\LB |\3D_{\rm far}|^2+ |\3B_{\rm far}|^2\RB
=\frac12|\3F_{\rm far}|^2
\end{align*}
and, by \eq{tem}, the far-zone Poynting vector is
\begin{align*}
\3S_{\rm far}&=\3E_{\rm far}\times\3H_{\rm far}
=\3D_{\rm far}\times\3B_{\rm far}
=\frac1{2i}{\3F}^*_{\rm far}\times\3F_{\rm far}\\
&=\frac12 {\3F}^*_{\rm far}\times(\3e_r\times\3F_{\rm far})
=\frac12 |\3F_{\rm far}|^2\3e_r=\5E_{\rm far}\3e_r 
\end{align*}
since $\3F_{\rm far}$ is orthogonal to $\3e_r$.

\section{Equivalent currents}

In principle, the scalar source generates the charge-current density by \eq{P} and \eq{J}. But this would involve not only the messy disentangling of the real and imaginary parts of $\3P=S\3\p$ (with both factors complex), but also dealing with the singular nature of  $S$. While $S$ is well-defined mathematically as a distribution \ci{K3}, it seems to be of little \it direct \rm value from a practical point of view. Since $S$ is supported on the branch cut $\5B$, one expects the electromagnetic sources to consist of a surface charge density $j_0$ and a surface current density $\3j$. But it turns out that these these surface sources are singular on the branch circle $\5C$, where $\s=0$. The essence of the problem can be understood from a careful analysis, given in \ci{K1a}, of a much simpler case, which we now recall.

\bf Example: \rm The analytically continued Coulomb field due to point charge of strength $Q=1$ is 
\begin{align}\lab{Coul}
\3C(\3r-i\3a)=-\grad\,\frac 1{4\p\s}=\frac{\3r-i\3a}{4\p\s^3}.
\end{align}
Newman \ci{N73} has shown that this can be identified with a \it real \rm electromagnetic field $(\3D,\3B)$ by
\begin{align}\lab{Coul2}
\3C=\3D+i\3B,
\end{align}
interpreted as the flat-spacetime (zero-mass) limit of the Maxwell field in the Kerr-Newman solution in general relativity \ci{N65}, which represents a spinning black hole of unit charge.\footnote{When $\3D$ in \eq{Coul2} is reinterpreted as a \it Newtonian force field, \rm then $\3B$ is a \it gravitomagnetic \rm field related to the `dragging' of Einsteinian spacetime in the vicinity of a spinning body. Evidently, this effect survives the flat-spacetime limit as the conjugate-harmonic partner to Newtonian gravitation.} It is instructive to compute the surface sources on a branch cut, which for simplicity we now take to be the disk $\5S_0$  defined in \eq{S0}. On the upper and lower faces of $\5S_0$ we have 
\begin{align*}
p\to +0, \ \   z\to\pm 0,\ \  \s\to \mp i\sr{a^2-\r^2},
\end{align*}
hence 
\begin{align*}
\3C\to\mp \frac{i\3\r}{4\p(a^2-\r^2)^{3/2}}
\mp\frac{\3a}{4\p(a^2-\r^2)^{3/2}}\,.
\end{align*}
The jumps in $\3D$ and $\3B$ across the cut are therefore
\begin{align*}
\d\3D=-\frac{\3a}{2\p(a^2-\r^2)^{3/2}}\,, \qq 
\d\3B=-\frac{2\3\r}{2\p(a^2-\r^2)^{3/2}}\,.
\end{align*}
Since $\d\3D$ is orthogonal and $\d\3B$ is tangent to  $\5S_0$, it follows that \it the  magnetic surface  charge- and current densities vanish as required. \rm The electric 
surface densities are given by \ci[p 18]{J99}
\begin{align}\lab{CoulSources}
j_0&=-\bh a\cdot\d\3D=-\frac{a}{2\p(a^2-\r^2)^{3/2}}\\
\3j&=\bh a\times\d\3B=-\frac{c\bh a\times\3\r}{2\p(a^2-\r^2)^{3/2}}
=-\frac{c\r\,\3e_{\f}}{2\p(a^2-\r^2)^{3/2}}\,,\nt
\end{align}
where we have inserted the speed of light (taken ealier to be $c=1$) for dimensional reasons. Before discussing the problem with \eq{CoulSources}, note that if we define the \it local charge velocity \rm by
\begin{align}\lab{vel}
\3v\equiv\frac{\3j}{\s}=\frac{c\r}a\,\3e_\f\,,
\end{align}
its linearity in $\r$  suggests a `hydrodynamic' interpretation of $\5S_0$ as a \it rigidly spinning charged disk with angular velocity \rm 
\begin{align}\lab{ang}
\O=c/a.
\end{align}
In particular,  \it the rim $\5C$ is moving at the speed of light. \rm 
While this conclusion seems bizarre in ordinary electrodynamics, it is entirely consistent with the origin of the field $\3C$ as the residual Maxwell field of a charged, spinning black hole.  The investigation in \ci{K1a} has sparked a renewed interest in Newman's original paper \ci{N73}, leading to similar interpretations of linearized gravitational fields \ci{N2} and a generzlized Lienard-Wiechert  field where the radiating point source moves along an arbitrary trajectory in \it complex \rm spacetime \ci{N4}. Our antennas will be similar, but their source is a \it dipole \rm following a complex trajectory and not a monopole, and so their charge-current densities are generated by \sl polarizations. \rm

We now come to the main lesson taught by this example. $\3C$ is an analytic continuation of the Coulomb field of a point source with charge $Q=1$. If the continuation is to make physical sense, the total charge should remain unchanged. This is contradicted by $ j_0$, which is not only strictly \it negative \rm but whose total charge on $\5S_0$  is $-\8$! To resolve this difficulty, it is necessary to treat the charge-current density as a singular  \it volume distribution, \rm just as the scalar source $S=\Box\Y$ was treated in \ci{K3}. The inhomogeneous Maxwell equations now state that the (volume) charge- and current density are
\begin{align}\lab{Jcoul}
J_0=\div\3F,\qq \3J=-\pl_t\3F-i\curl\3F,
\end{align}
while the \it homogeneous \rm Maxwell equations require that $J_0$ and $\3J$ be real. Taken as \it definitions \rm of the sources in the sense of generalized functions, it was shown in \ci{K3a} that \eq{Jcoul} indeed give a sensible answer. The equivalent surface sources on any spheroid $\5S_\a$ with $\a>0$ are defined by
\begin{align}\lab{BC}
j_0=\3e_p\cdot \d\3F,\qq \3j=-i\3e_p\times\d\3F,
\end{align}
where the outgoing unit normal $\3e_p$  on $\5S_\a$ is computed in the Appendix. These  sources are found to be  \it complex, \rm which means that they include a \it magnetic \rm charge-current density; the latter vanishes in the limit $\a\to 0$, in agreement with the above conclusion. The advantage of using $\a>0$ is that the sources are \it smooth and bounded, \rm with a total charge $Q=1$ as required. As $\a\to 0$, they decompose into \it surface  sources on the \it interior \rm of the disk which coincide with \eq{CoulSources}, plus \it line sources \rm on the rim $\5C$. The line sources carry a total charge of $\8$, but when the entire source distribution is treated as a generalized function, it carries the correct total charge  $Q=1$. The problem with \eq{CoulSources} is that the jump conditions (using infinitesimal pillboxes and loops) can be applied only on the interior of the disk and not on its boundary $\5C$. \it A similar argument applies to every branch cut, \rm  showing that caution must be exercised in computing equivalent sources,  a lesson we will recall when computing the currents required to produce electromagnetic wavelets.

Finally, we turn to computing the equivalent sources for $\3F$ on the spheroid $\5S_\a$. Some important properties of equivalent \it real scalar \rm surface sources were analyzed in \ci{HLK0}, but their connection to the vector case and, specifically, to our topological use of branch cuts, remains  to be explored.

According to \eq{joint} and \eq{Fo}, the jump in the field across the spheroid is
\begin{align}\lab{dF}\bx{\ 
\d\3F(\s,\t)=\3F(\s,\t)-\3F(-\s,\t)\ }
\end{align}
where the complex distance $\s$  with respect to $\5S_0$ is \it continuous \rm across $\5S_\a$. Unlike the \it sum \rm \eq{Fo} of retarded and advanced fields, the difference \eq{dF} \it does \rm have sources and they are confined to the surface $\5S_\a$, which we shall presently compute.  Begin by writing  \eq{F2} in the more explicit form
\begin{align}\lab{F3}
\3F(\s,\t)=L(\s,\t)\l\3u-M(\s,\t)\3\p-iN(\s,\t)\3u\times\3\p
\end{align}
with $L, M, N$ given in terms of the \it retarded \rm signal 
$g_r(\s,\t)=g(\t-\s)$ by
\begin{align*}
L(\s,\t) &=\frac{\ddot g_r}{\s}+\frac{3\dot g_r}{\s^2}+\frac{3g_r}{\s^3}\nt\\
M(\s,\t) &=\frac{\ddot g_r}{\s}+\frac{\dot g_r}{\s^2}+\frac{g_r}{\s^3}\nt\\
N(\s,\t) &=\frac{\ddot g_r}{\s}+\frac{\dot g_r}{\s^2}\,.
\end{align*}
Define the \it mixed \rm signals $g\6\pm$ by
\begin{align*}
g\6\pm(\s,\t)=g(\t-\s)\pm g(\t+\s)
\end{align*}
and note that
\begin{align*}
\s\to-\s\imp \3u=\frac{\3z}\s\to -\3u,\qq \l=\3u\cdot\3\p\to-\l.
\end{align*}
Then we obtain the following expression for the field discontinuity:
\begin{align}\lab{F4}
\d\3F=\2L(\s,\t)\l\3u-\2M(\s,\t)\3\p-i\2N(\s,\t)\3u\times\3\p,
\end{align}
where
\begin{align*}
\2L(\s,\t) &=\frac{\ddot g\6+}{\s}+\frac{3\dot g\6-}{\s^2}
+\frac{3g\6+}{\s^3}\\
\2M(\s,\t) &=\frac{\ddot g\6+}{\s}+\frac{\dot g\6-}{\s^2}
+\frac{g\6+}{\s^3}\nt\\
\2N(\s,\t) &=\frac{\ddot g\6-}{\s}+\frac{\dot g\6+}{\s^2}\,.
\end{align*}
Before going on to compute the currents, note that \eq{joint} can be modified so that 
the interior field is \it any \rm  source-free solution of Maxwell's equations, \ie
\begin{align*}
&\3F_\a=\begin{cases}
2\3F(\s,\t) & \text{in\ } V'\\  \3F_{\rm int}(\3r,t) & \text{in\ } V
\end{cases}\\
&\div\3F_{\rm int}=0,\qqq i\pl_t\3F_{\rm int}=\curl\3F_{\rm int}.
\end{align*}
The choice of an interior solution  other than $\3F_0(\s,\t)$ will of course modify the equivalent sources on $\5S_\a$. However, unless $\3F_{\rm int}$ fits into the spheroidal geometry, the resulting sources can be expected to be much more complicated and unnatural, and probably will not benefit from the `magic' of complex source points.  Probably the most general class of interior fields that \it do \rm fit the geometry consists of arbitrary \it multiples \rm of $\3F_0$, \ie 
\begin{align*}
\3F_{\rm int}(\3r,t)=\n\3F_0(\s,\t).
\end{align*}
Then \eq{dF} is replaced by
\begin{align}\lab{gendF}
\d \3F(\s,\t)=\m\3F(\s,\t)-\n\3F(-\s,\t), \qq  \m+\n=2,
\end{align}
and all computations below easily generalize to this case. However, \it only when $\m=\n=1$ can the radiating surface be interpreted as a combination of branch cuts! \rm I believe that this case is the most natural and expect it also to be the most useful. For this reason, only it will be treated here, although our results easily extend to the case \eq{gendF}.

I now compute or estimate the various inner and outer products needed in \eq{BC}. Free use will be made of the results derived in the Appendix, and there is no pretense of rigor.  I will  assume that
\begin{align*}
0<\a\ll a,
\end{align*} 
which means that the spheroid is rather \it flat. \rm By \eq{pq} and \eq{grad2},
\begin{align*}
\r\app \sr{a^2-q^2},\qq z=\a q,\qq
|\grad p|\app \frac a{|\s|}\,,\qq |\grad q|\app \frac\r{|\s|}.
\end{align*}
Thus
\begin{align}\lab{z}
\3u=\frac{\3r-i\3a}\s=\frac{\3\r+(z-ia)\bh a}\s\app \frac{\3\r-i\3a}\s\,.
\end{align}
Recall that $\3\p$ is orthogonal to $\3a$, so that
\begin{align*}
\3\p=\p_\r\3e_\r+\p_\f\3e_\f
\end{align*}
and thus
 \begin{align*}
\l=\3u\cdot\3\p=\frac{\3r\cdot\3\p}\s=\frac{\3\r\cdot\3\p}\s
=\frac{\r\p_\r}\s.
\end{align*}
From \eq{pqhat} in the Appendix, the outgoing unit normal on $\5S_\a$ is 
\begin{align}\lab{n}
\3e_p&=\frac{\grad p}{|\grad p|}= \frac{\a\3r+q\3a}{\sr{\a^2+q^2}\sr{\a^2+a^2}}\app
\frac{q\bh a}{\sr{\a^2+q^2}}=\frac{q}{|\s|}\,\bh a\,.
\end{align}
The $\a^2$ term has been retained in the denominator to control the singularity at the equator. (This is the main advantage of using $\5S_\a$ instead of $\5S_0$.) 
The approximation \eq{n} fails very near the equator $q=0$, where $\3e_p$ is far from parallel to $\3a$, but the analysis in \ci{HLK0} suggests that, for \it scalar \rm wavelets at least, the immediate vicinity of $q=0$ can be ignored. More precisely, it was shown that for time-harmonic driving signals of frequency $\o$, the \it effective aperture, \rm emitting most of the radiation, consists of the front surface of the disk $\5S_0$ parameterized by
\begin{align}\lab{effap}
k\inv\le q\le a,\ \  \ie\  \   \r^2\le a^2-1/k^2, \qq k=\o/c.
\end{align}
Of course, this has significance only if $ka>1$. Lower frequencies generate mostly a reactive field that swirls around the source region and is eventually 
reabsorbed.\footnote{This is consistent with the general fact that `DC components do not propagate.' It is also the basis of one of the close connections between electromagnetic wavelets and mathematical wavelet theory, since it amounts to an \it admissibility condition \rm on electromagnetic wavelets \ci[p 214]{K94}.}
Thus,  to obtain a high radiation efficiency, it is necessary  to use signals
$g\0\t$ with little low-frequency content, such as  linear combinations of  high-order derivatives of the Cauchy kernel \eq{C}.  (Of course, a careful repetition of the analysis needs to be made specifically for the electromagnetic case.) 

The inner products needed to find $j_0$ are
\begin{align*}
\3e_p\cdot\3u&=\3e_p\cdot(\grad p-i\grad q)=|\grad p|\app\frac a{|\s|}\\
\3e_p\cdot\3\p&\app\frac q{|\s|}\,\bh a\cdot \3\p= 0\\
\3e_p\cdot(\3u\times\3\p)&=\3\p\cdot(\3e_p\times\3u)
=\3\p\cdot(\3e_p\times(\grad p-i\grad q))\\
&=i |\grad q|\3\p\cdot(\3e_q\times\3e_p)
=i |\grad q|\3\p\cdot\3e_\f\app\frac{i\r}{|\s|}\, \p_\f.
\end{align*}
The outer products needed for $\3j$ are
\begin{align*}
\3e_p\times\3u&=i|\grad q|\3e_\f\app i\frac{\r}{|\s|}\,\3e_\f\\
\3e_p\times\3\p&\app \frac{q}{|\s|}\,\bh a\times\3\p
\app\frac {i\s}{|\s|}\lp\p_\r\3e_\f-\p_\f\bh\r\rp\\
\3e_p\times(\3u\times\3\p)&=(\3e_p\cdot\3\p)\3u-(\3e_p\cdot \3u)\3\p
\app-(\3e_p\cdot \3u)\3\p\app -\frac{a}{|\s|}\,\3\p.
\end{align*}
Using these in \eq{BC} gives the approximate surface charge density
\begin{align}\lab{j0app}
\bx{\ \s|\s|j_0\app \2L a\r\p_\r+\2N\s\r\p_\f \ }
\end{align}
and the approximate surface current density
\begin{align}\lab{japp}
\bx{\ \s|\s|\3j\app \lp \2L\r^2\p_\r-\2M\s^2\p_\r+\2N a\r\p_\f\rp\3e_\f
+\lp\2M\s^2\p_\f+\2N a\s\p_\r\rp\3e_\r.\ }
\end{align}
As expected from our example of the analytic Coulomb potential, the equivalent sources on a spheroid are \it complex, \rm indicating the presence of unrealizable magnetic charges.  Since the magnetic sources in that example vanished as $\a\to 0$, it is reasonable to hope that this will also be the case here. As the spheroid $\5S_\a$ with $0<\a\ll a$ is very flat, it may be possible to choose the \it phase \rm of the polarization vector $\3\p$ (representing the mixture of electric and magnetic dipoles) so as to minimize the magnetic sources over $\5S_\a$ excluding the vicinity of the rim $\5C$, and the latter region can be ignored for highly oscillatory driving signals as shown in \ci{HLK0}. This question will be addressed in detail elsewhere.

Finally, we compute the \it impulse response \rm of the antenna, \ie the sources $j_\m$ when the driving signal is the impulse  
\begin{align*}
g_0\0t=\d\0t\imp g\0\t=\frac 1{2i\p\t}=C_1\0\t.
\end{align*}
Notice that the \it real \rm point source version of the scalar wavelet \eq{Yr} is then the \it retarded propagator \rm for the wave equation,
\begin{align}\lab{Y0}
\Y\to\Y_0(r,t)=\frac{\d(t-r)} r,\qq \Box\Y_0=4\p\d(\3r,t),
\end{align}
where the precise relation between $\Y$ and $\Y_0$ is given in terms of complex-distance potential theory in \ci{K3}. The mixed signals are
\begin{align*}
g\6+&=\frac1{2i\p(\t-\s)}+\frac1{2i\p(\t+\s)}=\frac{\t}{i\p u}\,,
\hb{\ where\ } u=\t^2-\s^2\\
g\6-&=\frac1{2i\p(\t-\s)}-\frac1{2i\p(\t+\s)}=\frac{\s}{i\p u}
\end{align*}
and their time derivatives are
\begin{align*}
\dot g\6+&=-\frac{\t^2+\s^2}{i\p u^2}, 
&& \ddot g\6+=\frac{2\t^3+6\t\s^2}{i\p u^3}\\
\dot g\6-&=-\frac{2\s\t}{i\p u^2},
&&\ddot g\6-=\frac{6\t^2\s+2\s^3}{ i\p u^3}.
\end{align*}
This gives
\begin{align*}
\2L&=\frac{15\s^4\t-10\s^2\t^3+3\t^5}{i\p\s^3u^3}\\
\2M&= \frac{9\s^4\t-2\s^2\t^3+\t^5}{i\p\s^3u^3} \\
\2N&=\frac{3\s^4+6\s^2\t^2-\t^4}{i\p\s^2u^3},
\end{align*}
which can be substituted into \eq{j0app} and \eq{japp} to obtain the impulse response.

In view of the discussion following \eq{effap}, we are actually more interested in the system's response to the \it bandbass  signal \rm in \eq{C},
\begin{align*}
C_n\0\t=(i\pl_t)^{n-1} C_1\0\t
=\frac{(n-1)!}{2\p i^n\t^n}=(-\pl_b)^{n-1} C_1\0\t . 
\end{align*}
The induced surface source  $j_\m^{\0n}$ can be computed directly from the impulse response:
\begin{align}
j_\m^{\0n}=(-\pl_b)^{n-1} j_\m.
\end{align}

\section{Concluding note}

Source-free relativistic fields always extend analytically to the double tube domain $\5T\6\pm$ \eq{Tpm} of complex space-time, as explained in \ci{K3}.  I find it quite remarkable that the extension $\Y(\s,\t)$ of the propagator \eq{Y0} generates fields with spatially compact sources that  are analytic in the source-free parts $\5T\6{\5B}$ of complex space-time obtained by removing 
 the \it world tubes \rm swept out by the sources. The boundary values of these analytic fields then characterize the singular sources, as shown above.

\section{Appendix}

The complex unit vector $\3u$ is given by
\begin{align}\lab{u}
\3u=\grad \s=\frac{\3z}{\s}=\grad p-i\grad q,
\end{align}
hence
\begin{align}\lab{grad}
\grad p=\frac{p\3r+q\3a}{p^2+q^2},\qq	 
\grad q=\frac{p\3a-q\3r}{p^2+q^2}.
\end{align}
Note that
\begin{align}\lab{orth}
\3u\cdot\3u=1\imp |\grad p|^2-|\grad q|^2=1,\qq \grad p\cdot\grad q=0
\end{align}
and 
\begin{align*}
|\grad p|^2+|\grad q|^2=
\3u^*\cdot\3u=\frac{r^2+a^2}{p^2+q^2}=\frac{p^2-q^2+2a^2}{p+2+q^2},
\end{align*}
which gives
\begin{align}\lab{grad2}
 |\grad p|^2=\frac{p^2+a^2}{p^2+q^2}, \qq  |\grad q|^2=\frac{a^2-q^2}{p^2+q^2}.
\end{align}
The unit vectors in the directions of increasing $p$ and $q$ are therefore
\begin{align}\lab{pqhat}
\3e_p&=\frac{p\3r+q\3a}{\sr{p^2+q^2}\sr{p^2+a^2}}\\
\3e_q&=\frac{p\3a-q\3r}{\sr{p^2+q^2}\sr{a^2-q^2}}. \nt
\end{align}

\section*{Acknowledgements}
It is a pleasure to thank Drs.~Richard Albanese,  Iwo Bialynicki-Birula, Ehud Heyman, Ted Newman, Ivor Robinson, Andzej Trautman and Arthur Yaghjian for friendly discussions and suggestions related to this work, and David Park for generous help with the figures using his DrawGrahics package.  I am especially grateful to Dr.~Arje Nachman for his sustained support of my research, most recently through AFOSR Grant  \#F49620-01-1-0271.

\end{document}